\documentclass[12pt,a4paper,english]{article}
\usepackage[authoryear]{natbib}
\usepackage{graphicx}
\usepackage{geometry}
\newcommand{\degr}{$^{\circ}$}  
\newcommand{\ignore}[1]{}
\newcommand{\ex}{$\times $\,} 

\makeatother
\begin{document}

\title{{A review of wildland fire spread modelling, 1990-present}\\
      { 3: Mathematical analogues and simulation models}}

\author{A.L. Sullivan}

\maketitle

{\small
\begin{center}

Ensis\footnote{A CSIRO/Scion Joint Venture} Bushfire Research \footnote{current address: Department of Theoretical Physics,\\
Research School of Physical Sciences and Engineering\\
The Australian National University, Canberra 0200, Australia. }\\
PO Box E4008, Kingston, ACT 2604, Australia

email: Andrew.Sullivan@ensisjv.com or als105@rsphysse.anu.edu.au

phone: +61 2 6125 1693, fax: +61 2 6125 4676

\end{center}}

\begin{center}version 3.0 \end{center}

\setlength\parskip{\bigskipamount} \setlength\parindent{0pt}

\begin{abstract}
In recent years, advances in computational power and spatial data analysis
(GIS, remote sensing, etc) have led to an increase in attempts to model the
spread and behvaiour of wildland fires across the landscape. This series of
review papers endeavours to critically and comprehensively review all types of
surface fire spread models developed since 1990. This paper reviews models of a
simulation or mathematical analogue nature. Most simulation models are
implementations of existing empirical or quasi-empirical models and their
primary function is to convert these generally one dimensional models to two
dimensions and then propagate a fire perimeter across a modelled landscape.
Mathematical analogue models are those that are based on some mathematical
conceit (rather than a physical representation of fire spread) that
coincidentally simulates the spread of fire. Other papers in the series review
models of an physical or quasi-physical nature and empirical or quasi-empirical
nature. Many models are extensions or refinements of models developed before
1990. Where this is the case, these models are also discussed but much less
comprehensively.
\end{abstract}


\section{Introduction}

\subsection{History}

The ultimate aim of any prediction system is to enable an end user to carry out
useful predictions.  A useful prediction is one that helps the user achieve a
particular aim.  In the field of wildland fire behaviour, that aim is primarily
to stop the spread of the fire or to at least reduce its impact on life and
property.  The earliest efforts at wildland fire behaviour prediction
concentrated on predicting the likely danger posed by a particular fire or set
of conditions prior to the outbreak of a fire.  These fire danger systems were
used to set the level of preparedness of suppression resources or to aid in the
identification of the onset of bad fire weather for the purpose of calling
total bans on intentionally lit fires.

In addition to a subjective index of fire danger, many of early fire danger
systems also provided a prediction of the likely spread of a fire, as a
prediction of the rate of forward spread of the fire, the rate of perimeter
increase or rate of area increase. In many cases, these predictions were used
by users to plot the likely spread of the fire on a map, thereby putting the
prediction in context with geographic features or resource locations, and
constituted the first form of fire spread simulation.

Because much of the development of the early wildland fire behaviour models was
carried out by those organisations intended to use the systems, the level of
sophistication of the model tended to match the level of sophistication of the
technology used to implement it.  Thus, the early fire spread models provided
only a single dimension prediction (generally the forward rate of spread of the
headfire) which could be easily plotted on a map and extrapolated over time.
While modern wildland fire spread modelling has expanded to include physical
approaches \citep{Sullivan2007b}, all modern operational fire spread models
have continued this empirical approach in the form of one-dimensional spread
models \citep{Sullivan2007c}.  Much of the development of technology for
implementing the models in a simulation environment has concentrated on methods
for converting the one-dimensional linear model of fire spread to that of
two-dimensional planar models of fire spread.

In parallel with approaches to implement existing empirical models of fire
spread have been efforts to approach the simulation of fire spread across the
landscape from a holistic perspective. This has resulted in the use of methods
other than those directly related to the observation, measurement and modelling
of fire behaviour.  These methods are mathematical in nature and provide an
analogue of fire behaviour.  Many of these approaches have also paralleled the
development of the computer as a computational device to undertake the
calculations required to implement the mathematical concepts.

An increase in the capabilities of remote sensing, geographical information
systems and computing power during the 1990s resulted in a revival in the
interest of fire behaviour modelling as applied to the prediction of spread
across the landscape.

\subsection{Background}

This series of review papers endeavours to comprehensively and critically
review the extensive range of modelling work that has been conducted in recent
years. The range of methods that have been undertaken over the years represents
a continuous spectrum of possible modelling \citep{Karplus1977a}, ranging from
the purely physical (those that are based on fundamental understanding of the
physics and chemistry involved in the behaviour of a wildland fire) through to
the purely empirical (those that have been based on phenomenological
description or statistical regression of fire behaviour). In between is a
continuous meld of approaches from one end of the spectrum or the other.
\citet{Weber1991a} in his comprehensive review of physical wildland fire
modelling proposed a system by which models were described as physical,
empirical or statistical, depending on whether they account for different modes
of heat transfer, make no distinction between different heat transfer modes, or
involve no physics at all. \citet{Pastor2003} proposed descriptions of
theoretical, empirical and semi-empirical, again depending on whether the model
was based on purely physical understanding, of a statistical nature with no
physical understanding, or a combination of both. \citet{Grishin1997} divided
models into two classes, deterministic or stochastic-statistical. However,
these schemes are rather limited given the combination of possible approaches
and, given that describing a model as semi-empirical or semi-physical is a
`glass half-full or half-empty' subjective issue, a more comprehensive and
complete convection was required.

Thus, this review series is divided into three broad categories: Physical and
quasi-physical models; Empirical and quasi-empirical models; and Simulation and
Mathematical analogous models. In this context, a physical model is one that
attempts to represent both the physics and chemistry of fire spread; a
quasi-physical model attempts to represent only the physics. An empirical model
is one that contains no physical basis at all (generally only statistical in
nature), a quasi-empirical model is one that uses some form of physical
framework upon which to base the statistical modelling chosen. Empirical models
are further subdivided into field-based and laboratory-based. Simulation models
are those that implement the preceding types of models in a simulation rather
than modelling context. Mathematical analogous models are those that utilise a
mathematical precept rather than a physical one for the modelling of the spread
of wildland fire.

Since 1990, there has been rapid development in the field of spatial data
analysis, e.g. geographic information systems and remote sensing. As a result,
I have limited this review to works published since 1990. However, as much of
the work that will be discussed derives or continues from work carried out
prior to 1990, such work will be included much less comprehensively in order to
provide context.

\subsection{Previous reviews}

Many of the reviews that have been published in recent years have been for
audiences other than wildland fire researchers and conducted by people without
an established background in the field. Indeed, many of the reviews read like
purchase notes by people shopping around for the best fire spread model to
implement in their part of the world for their particular purpose. Recent
reviews (e.g. \citet{Perry1998,Pastor2003}; etc), while endeavouring to be
comprehensive, have offered only superficial and cursory inspections of the
models presented. \citet{Morvan2004a} takes a different line by analysing a
much broader spectrum of models in some detail and concludes that no single
approach is going to be suitable for all uses.

While the recent reviews provide an overview of the models and approaches that
have been undertaken around the world, mention must be made of significant
reviews published much earlier that discussed the processes in wildland fire
propagation themselves. Foremost is the work of \citet{Williams1982}, which
comprehensively covers the phenomenology of both wildland and urban fire, the
physics and chemistry of combustion, and is recommended reading for the
beginner. The earlier work of \citet{Emmons1963,Emmons1966} and \citet{Lee1972}
provides a sound background on the advances made during the post-war boom era.
\citet{Grishin1997} provides an extensive review of the work conducted in
Russia in the 1970s, 80s and 90s.

The first paper in this series discussed those models based upon the
fundamental principles of the physics and chemistry of wildland fire behaviour.
The second paper in the series discussed those models based directly upon only
statistical analysis of fire behaviour observations or models that utilise some
form of physical framework upon which the statistical analysis of observations
have been based.  Particular distinction was made between observations of the
behaviour of fires in the strictly controlled and artificial conditions of the
laboratory and those observed in the field under more naturally occurring
conditions.

This paper, the final in the series, focuses upon models concerned only with
the simulation of fire spread over the landscape and models that utilise
mathematical conceits analogous to fire spread but which have no real-world
connection to fire.  The former generally utilise a pre-existing fire spread
model (which can be physical, quasi-physical, quasi-empirical or empirical) and
implements it in such a way as to simulate the spread of fire across a
landscape.  As such, it is generally based upon a geographic information system
(GIS) of some description to represent the landscape and uses a propagation
algorithm to spread the fire perimeter across it.  The latter models are for
the most part based upon accepted mathematical functions or concepts that have
been applied to wildland fire spread but are not derived from any understanding
of wildland fire behaviour.  Rather, these models utilise apparent similarities
between wildland fire behaviour and the behaviour of these concepts within
certain limited contexts.  Because of this, these mathematical concepts could
equally be applied to other fields of endeavour and, for the most part have
been, to greater or lesser success.

Unlike the preceding entries in this series, this paper is segmented by the
approaches taken by the various authors, not by the authors or their
organisations, given the broad range of authors that in some instances have
taken similar approaches.

\section{Fire Spread Simulations}

The ultimate aim of any fire spread simulation development is to produce a
product that is practical, easy to implement and provides timely information on
the progress of fire spread for wildland fire authorities.  With the advent of
cheap personal computing and the increased use of geographic information
systems, the late 1980s and early 1990s saw a flourishing of methods to predict
the spread of fires across the landscape \citep{Beer1990a}.  As the generally
accepted methods of predicting the behaviour of wildland fires at that time
were (and still are) one-dimensional models derived from empirical studies
\citep{Sullivan2007b}, it was necessary to develop a method of converting the
single dimension forward spread model into one that could spread the entire
perimeter in two dimensions across a landscape. This involves two distinct
processes: firstly, representing the fire in a manner suitable for simulation,
and secondly, propagating that perimeter in a manner suitable for the
perimeter's representation.

Two approaches for the representation of the fire have been implemented in a
number of softwares. The first treats the fire as a group of mainly contiguous
independent cells that grows in number, described in the literature as a raster
implementation. The second treats the fire perimeter as a closed curve of
linked points, described in the literature as a vector implementation.

The propagation of the fire is then carried using some form of expansion
algorithm. There are two main methods used. The first expands the perimeter
based on a direct-contact or near-neighbour proximity spread basis. The second
is based upon Huygens' wavelet principle in which each point on the perimeter
becomes the source of the interval of fire spread. While the method of
propagation and method of fire representation are often tied (for example,
Hyuygens' wavelet principle is most commonly used in conjuction with a vector
representation of the fire perimeter, there is no reason why this should be so
and methods of representation and propagation can be mixed.

\subsection{Huygens wavelet principle}

Huygen's wavelet principle, originally proposed for the propagation of light
waves, was first proposed in the context of fire perimeter propagation by
\cite{Anderson1982}.  In this case, each point on a fire perimeter is
considered a theoretical source of a new fire, the characteristics of which are
based upon the given fire spread model and the prevailing conditions at the
location of the origin of the new fire. The new fires around the perimeter are
assumed to ignite simultaneously, to not interact and to spread for a given
time, $\Delta t$.  During this period, each new fire attains a certain size and
shape, and the outer surface of all the individual fires becomes the new fire
perimeter for that time.

\begin{figure}[p!]
  \includegraphics[width=10cm]{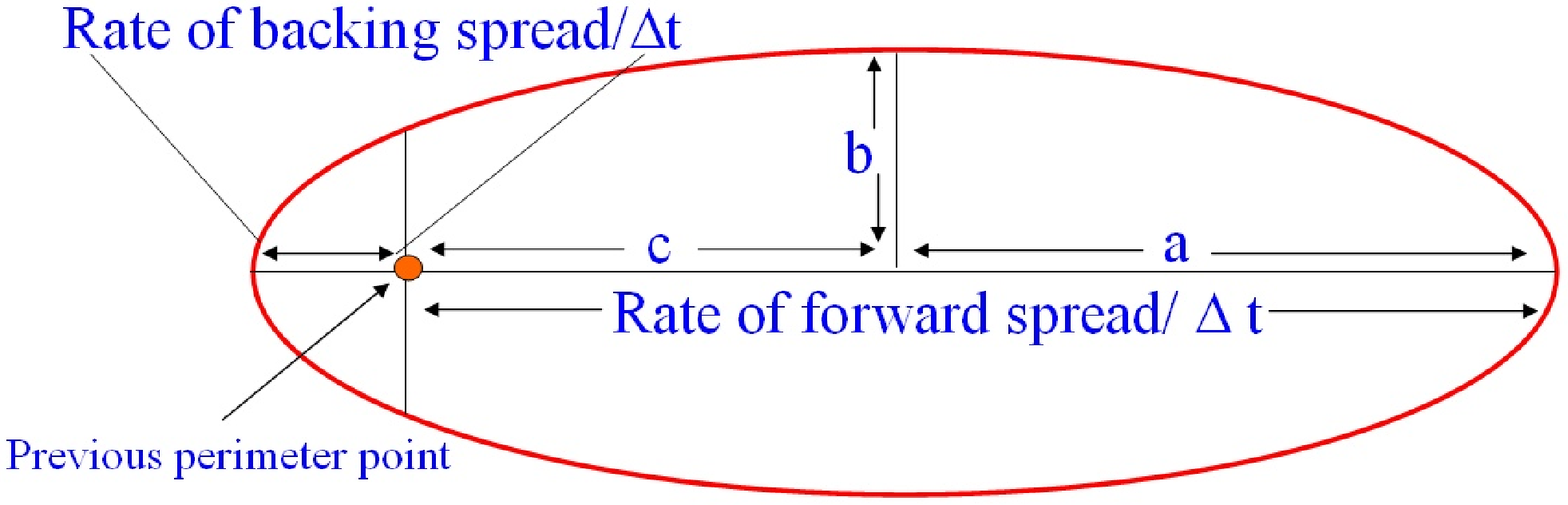}\\
  \caption{Ellipse geometry determined from the fire spread and length:breadth model.
  a+c determined from the predicted rate of spread, b is determined from a and length:breadth
  model.}\label{Ellipse}
\end{figure}

\cite{Anderson1982} used an ellipse to define the shape of the new fires with
the long axis aligned in the direction of the wind. Ellipse shapes have been
used to described fire spread in a number of fuels
\citep{Peet1965,McArthur1966,VanWagner1969} and, although many alternative and
more complex shapes have been proposed (e.g. double ellipse, lemniscate and
tear drops \citep{Richards1995a}, the ellipse shape has been found to
adequately described the propagation of wildland fires allowed to burn
unhindered for considerable time \citep{Anderson1982,Alexander1985}. The
geometry of the ellipse template is determined by the rate of forward spread as
predicted by the chosen fire spread model and a suitable length-to-breadth
ratio (L:B) to give the dimensions of the ellipse (Fig. \ref{Ellipse}).
\citet{McArthur1966} proposed ratios for fires burning in grass fuels in winds
up to $\simeq$ 50 km h$^{-1}$. \citet{Alexander1985} did the same for fires in
conifer forests also up to a wind speed of $\simeq$ 50 km h$^{-1}$.

\begin{figure}[p!]
  \includegraphics[width=15cm]{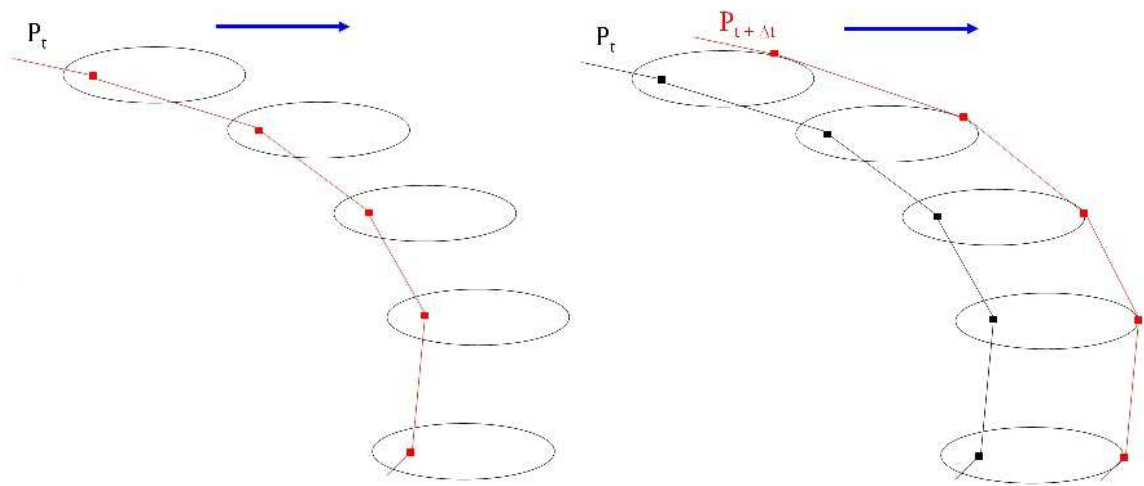}\\
  \caption{A schematic of the application of Huygens wavelet principle to
  fire perimeter propagation. In the simple case of homogeneous fuel, a uniform
  template ellipse, whose geometry is defined by the chosen fire spread
  model, length:breadth ratio model and the given period of propagation
$\Delta t$,  is applied to each node representing the current perimeter.  The
new perimeter at $t + \Delta t$ is defined by the external points of all new
ellipses.}\label{Huygens}
\end{figure}

Figure \ref{Huygens} illustrates the application of Huygens' principle to the
propagation of a fire perimeter utilising the ellipse template. A section of
perimeter defined by a series of linked nodes that act as the source of a
series of new fires. The geometry of the ellipse used for each new fire is
determined by the prevailing conditions, the chosen fire spread model and
length-breadth ratio model, and the given period of propagation $\Delta t$.  In
the simple case of homogeneous conditions, all ellipses are the same and the
propagation is uniform in the direction of the wind. The boundary of the new
ellipses forms the new perimeter at $t + \Delta t$.

\citet{Richards1990, Richards1995a} produced analytical solutions for this
modelling approach, for a variety of template shapes, in the form of a pair of
differential equations. A computer algorithm \cite{Richards1995b} that utilises
these equations was developed and was subsequently incorporated into fire
simulation packages, including FARSITE (USA) \citep{Finney1994, Finney1998} and
Prometheus (Canada) \citep{CWFGM2004}.  An alternative method that utilises the
elliptical geometry only is that of \citet{Knight1993} which is used in
SiroFire (Australia) \citep{Beer1990a, Coleman1996}.  This method provides
solutions to the two main problems with the closed curve expansion approach
using Huygens' wavelet principle, namely rotations in the perimeter, in which a
section of perimeter turns itself inside-out, and enclosures, in which unburnt
fuel is enclosed by two sections of the same perimeter (termed a `bear hug'). A
similar method was proposed by \citet{Wallace1993}.

FARSITE is widely used in the US by federal and state land management agencies
for predicting fire spread across the landscape. It is based upon the BEHAVE
\citep{Andrews1986} fire behaviour prediction systems, which itself is based
upon the spread model of \citet{Rothermel1972}. It includes models for fuel
moisture content \citep{Nelson2000}, spotting \citep{Albini1979a}, post-front
fuel consumption \citep{Albini1995a, Albini1995b}, crown-fire initiation
\citep{vanWagner1977b} and crown-fire spread \citep{Rothermel1991}. It is
PC-based in MS-Windows and utilises the ARCView GIS system for describing the
spatial fuel data and topography.

SiroFire was developed for operational use in Australia and utilises McArthur's
fire spread models for grass \citep{McArthur1966} and forest
\citep{McArthur1967} as well as the recommended replacement grassland model
\citep{Cheney1998} and versions of Rothermel's models configured for Australian
grass and forest litter fuel, however while it was never used operationally it
did find use as a training tool for volunteer bushfire firefighters.  It uses a
proprietary geographic format intended to reduce computation time with data
derived from a number of GIS platforms. It was PC-based using DOS protected
mode although would run under MS-Windows.  It has now been subsumed into a risk
management model, Phoenix, being developed by the University of Melbourne

The Canadian Wildland Fire Growth Model, Prometheus, is based on the Canadian
Fire Behaviour Prediction (FBP) System \citep{FCFDG1992} and utilises the
wavelet propagation algorithms of \citet{Richards1995a, Richards1995b} to
simulate the spread of wildland fire across landscapes. It was initially
developed by the Alberta Sustainable Resource Development, Forest Protection
Division for the Canadian Forest Service, but is now a national interagency
endeavour across Canada endorsed by the Canadian Interagency Forest Fire
Centre. It is Windows-based, utilises maps and geographic data exported from
the Esri GIS platform ARC and is intended for use as a realtime operational
management tool. As with FARSITE and SiroFire, Prometheus allows the user to
enter and edit fuel and meteorological data and carry out simulations of fire
spread.

The symmetric nature of the template ellipse in conjunction with the
application of Huygens' wavelet principle neatly provides the flank and rear
spread of a fire.  By relating the flank and rear spread through the ellipse
geometry, the single forward rate of spread of the fire is all that is needed
to simulate the spread of the entire perimeter.  \citet{French1990} found that
in homogeneous fuels and weather conditions, the Huygens' wavelet principle
with template ellipse shape suitably modelled fire spread, with only small
distortion of the fire shape. However, such a method cannot adequately handle
conditions and fuels that are heterogeneous and errors introduced through
changes in the conditions during the period $\Delta t$ as well as distortions
in the fire perimeter due to artifacts in the Huygens' wavelet method.

Changes in conditions during the propagation period $\Delta t$ cause the
predicted perimeter to over- or under-predict the spread of the perimeter
because those changes are not reflected in the predicted perimeter.  Reducing
$\Delta t$ can reduce the impact of such changes and a flexible approach to the
setting of $\Delta t$ has been used with great success \citep{Finney1998}.

\citet{Eklund2001} implemented the method of \citet{Knight1993} as a fire
propagation engine existing with a geographic database on a distributed network
such as the World Wide Web (WWW).

\subsection{Raster-based simulation}

In a raster-based simulation, the fire is represented by a raster grid of cells
whose state is unburnt, burning or burnt.  This method is computationally less
intensive than that of the closed curve (vector) approach, and is much more
suited to heterogenous fuel and weather conditions.  However, because fuel
information needs to be stored for each and every cell in the landscape, there
is a trade-off between the resolution at which the data is stored and the
amount of data that needs to be stored (and thus memory requirements and access
times, etc)\footnote{In vector data, fuel is stored as polygons represented by
a series of data points representing the vertices of the outline of the fuel
and the fuel attributes for the whole polygon.  Very large areas can be stored
in this fashion but with overhead in processing to determine if a point is
inside the polygon.}.

The method of expanding the fire in this fashion is similar to that of cellular
automata, in which the fire propagation is considered to be a simple set of
rules defining the interaction of neighbouring cells in a lattice.  I will
differentiate fire propagation simulations that utilise a pre-existing fire
spread model to determine the rate of fire expansion from those that are true
cellular automata.  The former are described here as raster- or grid-based
simulations and the latter are dealt with in the following section on
mathematical analogues.

\citet{Kourtz1971} were the first to apply computer techniques to the modelling
of fire spread across a landscape.  Initially simulating the smouldering spread
of small fires ($<0.02$ ha) using a grid of 50 \ex 50 square cells each of 1
ft$^2$ in no wind and no slope, this model was extended using a combination of
Canadian and US \citep{Rothermel1972} fire behaviour models \citep{Kourtz1977}
and the output was in the form of text-based graphical representation of
predicted spread. \cite{King1971} developed a model of rate-of-area increase of
aerial prescribed burns (intended for use on a hand-held calculator) based on
an idealised model of the growth of a single spot fire. \citet{Fransden1979}
utilised a hexagonal lattice to represent heterogeneous fuel beds and a
least-time-to-ignition heat balance model to simulate fire spread across it.
\citet{Green1983a,Green1983b} generalised the approach of Kourtz and O'Regan
and investigated the effect of discontinuous, heterogeneous fuel in square
lattices on fire shape utilising both a heat balance and a flame contact spread
models and found that while fire shapes are less regular than in continuous
fuels, the fires tended to become more elliptical in shape as the fire
progressed, regardless of the template shape used.

\citet{Green1990} produced a landscape modelling system called IGNITE that
utilised the fire spread mechanics of \citet{Green1983a}. This system is a
raster-based fire spread model that uses fire spread models of
\citet{McArthur1966, McArthur1967} as retro-engineered by \citet{Noble1980} and
an elliptical ignition template to predict the rate of forward spread in the
form of ``time to ignition'' for each cell around a burning cell. IGNITE very
easily deals with heterogenous fuels and allows the simulation of fire
suppression actions through changes in the combustion characteristics of the
fuel layers.

\citet{Kalab1991} and \citet{Vascon1992} introduced similar methods to
spatially resolve Rothermel's spread model in BEHAVE by linking it to
raster-based GIS platforms. \citet{Kalab1991} developed a simulation technique
that derived a `friction' layer within the GIS for six base spread rates for
which the friction value increased as spread rate decreased.  This was combined
with six wind speed classes to produce a map of potential fire extent contours
and fireline intensity strata across a range of slope and aspect classes.
\citet{Vascon1992} developed a similar simulation package called FIREMAP that
continued the earlier work of \citet{Vascon1990}. FIREMAP stored topographic,
fuel and weather information as rasterised layers within the GIS. It is assumed
that the resolution of the rasters are such that all attributes within each
cell are uniform. Fire characteristics such as rate of spread, intensity,
direction of maximum spread, flame length, are calculated for each cell and
each weather condition to produce a database of output maps of fire behaviour.
Simulation is then undertaken by calculating each cell's `friction' or time
taken for a fire front to consume a cell. \citet{Ball1992} extended the work of
Vasconcelos by improving the method used to implement the cell to cell spread
by adjusting the ROS for flank and rear spread based upon BEHAVE's cosine
relation with head fire ROS. The authors found the resulting predicted fire
shapes to be unnaturally angular and attribute this to the poor relation for
flank spread given by BEHAVE, the regular lattice shape of the raster, and the
fact that spread angles are limited, concluding that the raster structure
cannot properly represent 'the continuous nature of the actual fire'.

\citet{Karafyllidis1997} developed a raster-based simulation also based on
\cite{Rothermel1972} for hypothetical landscapes. The state of each raster cell
is the ratio of the area burned of the cell to the total area of the cell. The
passage of the fire front is determined by the sum of the states of each cell's
neighbours at each time step until the cell is completely burnt. This approach
requires, as input parameter for each cell, the rate of spread of a fire in
that cell based on the fuel alone.  \citet{Berjak2002} improved the model by
incorporating the effects of slope and wind on the scalar field of cell rate of
spread using the slope effect model of \citet{Cheney1981} and an empirical
flame angle/wind speed function.  This model was then applied to spatially
heterogeneous Savanna fuels of South Africa and found to be in good agreement
with observed fire spread.

FireStation \citep{Lopes1998,Lopes2002} and PYROCART \citep{Perry1999} both
implement Rothermel's fire spread model in a raster-based GIS platform.
FireStation utilises both single and double-ellipse fire shape templates,
depending on wind speed, to dictate the spread across cells. The 3-dimensional
wind field across the landscape is based on local point observations
extrapolated using either a linear conservation of mass approach, or a full
3-dimensional solution of the Navier-Stokes equations. Slope is treated as an
`equivalent' wind. PYROCART utilises the fire shape model of \citet{Green1990}
which is a function of wind speed. It was validated against a small wildfire
and its predictive accuracy (a measure of performance based on the percentage
of cells predicted to be burnt compared to those that were unburnt or not
predicted to burn) estimated to be 80\%.

\citet{Guariso2002} extend the standard 2-dimensional approach to modelling the
spread of a surface fire by implementing two levels of raster-based models, one
to represent surface fuel and its combustion and another to represent, once a
critical threshold value has been reached, the forest canopy and its
combustion. They utilise Rothermel's fire spread model with fuel models
modified and validated for Mediterranean fuel complexes.  To improve its
capabilities as an operational tool, fire fighting resources are tracked on
screen using Global Positional System (GPS). \citet{Trunfio2004} implemented
Rothermel's model using a hexagonal cell shape and found that the model did not
produce the spurious symmetries found with square-shaped lattices.

\subsection{Other propagation methods}

There are alternatives to the raster cell or vector ellipse template
propagation methods described above, although these are less widespread in
their use. Coupled fire-atmosphere models that incorporate a pre-existing fire
spread model (such as given by Rothermel or McArthur) are, at their most basic,
a form of propagation algorithm. The coupled fire-atmosphere model of
\citet{Clark1996a,Clark1996b,Clark2004} represents a considerable effort to
link a sophisticated 3-dimensional high-resolution, non-hydrostatic mesoscale
meteorological model to a fire spread model.  In this particular case, the
mesoscale meteorological model was originally developed for modelling
atmospheric flows over complex terrain, solving the Navier-Stokes and
continuity equations and includes terrain following coordinates, variable grid
size, two-way interactive nesting, cloud (rain and ice) physics, and solar
heating \citep{Coen2005}.  It was originally linked to the empirical model of
forest fire spread of \citet{McArthur1967} \citep{Clark1998a} but was later
revised to incorporate the spread model of Rothermel instead \citep{Coen2000}.

The atmosphere model is coupled to the fire spread model through the sensible
(convection and conduction effects) and latent heat fluxes approximated from
the fireline intensity (obtained via the ROS) predicted by the model for a
given fuel specified in the model. Fuel is modelled on a raster grid of size 30
m \citep{Coen2005}. Fuel moisture is allowed to vary diurnally following a very
simple sinusoid function based around an average daily value with a fixed lag
time behind clock time \citep{Coen2005}. Assumptions are made about the amount
of moisture evaporated prior to combustion. Fuel consumption is modelled using
the BURNUP algorithm of \citet{Albini1995a}. Effects of radiation, convection
and turbulent mixing occurring on unresolved scales (i.e. $<$ 30 m) are
`treated crudely' without any further discussion.

The coarse nature of the rasterised fuel layer meant that a simple cell fire
spread propagation technique was too reliant on the cell resolution. A fire
perimeter propagation technique that is a unique mix of the raster- and the
vector-based techniques was developed. Each cell in the fuel layer is allowed
to burn at an independent rate, dependent upon the predicted wind speed at a
height of 5 m, the predicted rate of spread and the fuel consumption rate. Four
tracers aligned with the coordinate system, each with the appropriate ROS in
the appropriate directions (headfire ROS is defined as that parallel to the
wind direction) are used to track the spread of fire across a fuel cell. The
coordinates of the tracers define a quadrilateral that occupies a fuel cell
which is allowed to spread across the fuel cell. The tracers move across a fuel
cell until they reach a cell boundary. If the adjacent cell is unburnt, it is
ignited and a fresh set of tracers commenced for the boundaries of that cell.
Meanwhile, once the tracers reach a cell boundary, they can then only move in
the orthogonal direction. In this way, the quadrilateral can progress across
the cell. The boundaries of all the quadrilaterals then make up the fire
perimeter. The size of the quadrilateral then allows an estimate of the amount
of fuel that has been consumed since the cell ignited. The fireline propagation
method allows for internal fire perimeters, although it only allows one
fireline per fuel cell.

The interaction between the heat output of the burning fuel and the
3-dimensional wind field results in complex wind patterns, which can include
horizontal and vertical vortices. \citet{Clark1996a, Clark1996b, Jenkins2001b}
explored these in producing fireline phenomena such as parabolic headfire
shape, and convective and dynamic fingering. However, as wind speed increases
($>$ 5 m s$^{-1}$ at 15 m above ground), \citet{Clark1996b} found that the
coupling weakens and the wind flows through the fire.

The real utility of the coupled fire-atmosphere model, however, is the
prediction of wind direction around the fire perimeter, used to drive the
spread of the fire. This, in effect, replaces Huygens' wavelet approach with a
much more physically direct method. However, the use of Rothermel's fire spread
model for spread in directions other than in the direction of the prevailing
wind is questionable and results in odd deformations in the fire perimeter when
terrain or fuel are not uniform \citep{Clark2004}.

Several other workers have taken the same approach as Clark in linking a
mesoscale meteorological model to a fire model. \citet{Gurer1998} coupled an
off-the-shelf mesoscale meteorological model (the Regional Atmospheric Modeling
System (RAMS) \citep{Pielke1992} with Rothermel's model of fire spread to
predict gas and particulate fall out from forest fires for the purpose of
safety and health. The Rothermel model is used to obtain burning area and heat
for input into RAMS. Submodels are used for prediction of the emission
components (CO2, CH4, etc, polycyclic aromatic hydrocarbons, etc). Simulation
of the fire perimeter propagation is not undertaken. \citet{Speer2001} used a
numerical weather prediction model to predict the speed and direction of the
wind for input into a simple empirical model of fire spread through heathland
fuel to predict the rate of forward spread (not to simulate the spread) of two
wildfires in Sydney 1994.

\citet{Plourde1997} extend the application of Huygens's wavelet propagation
 principle as utilised by Knight and Coleman (1993). However, rather than
relying on the template ellipse as the format for the next interval
propagation, the authors utilise an innovative closed contour method based on a
complex Fourier series function. Rather than considering the perimeter as a
series of linked points that are individually propagated, the perimeter is
considered as a closed continuous curve that is propagated in its entirety. A
parametric description of the perimeter is derived in which the x and y
coordinates of each point are encoded as a real and imaginary pair. However, as
with \citet{Knight1993}, a sufficiently fine time step is critical to precision
and anomalies such as rotations and overlaps must be identified and removed.
\citeauthor['s]{Plourde1997} propagation model appears to handle heterogeneous
fuel but the timestep is given as 0.05 s, resulting in very fine scale spread
but with the trade-off of heavy computational requirements.

\citet{Viegas2002} proposed an unorthodox propagation mechanism in which the
fire perimeter is assumed to be a continuous entity that will endeavor to
rotate to align itself with a slope to an angle of approximately 60 degrees
across the slope. Based on observation of laboratory and field experiments of
line fires lit at angles to the slope, Viegas constructs a fire perimeter
propagation algorithm in which he redefines the flank spread of a fire burning
in a cross-slope wind as a rotation of the front.

In perhaps a sign of the times, \citet{Seron2005} takes the physical model of
\citet{Asensio2002} and simulates it using the techniques and approaches
developed for computer generated imagery (CGI). A strictly non-realtime method
is used to solve the fire spread model utilising satellite imagery and the
terrain (using flow analysis techniques), interpolated using kriging, to
determine fuel and non-burnables such water bodies and rivers, etc. All
attributes are non-dimensionalised. Wind is calculated as a vector for each
cell derived from the convection form of the physical fire model
\citep{Asensio2005}. This vector is then added to the terrain gradient vector.
256 x 256 cells are simulated, resulting in 131589 equations that need to be
solved for each timestep, which is 0.0001 s.

\section{Mathematical Analogues}

In the broader non-wildland-fire-specific literature there is a considerable
number of works published involving wildland fire spread that are not based on
wildland fire behaviour. For the most part, these works implement mathematical
functions that appear analogous to the spread of fires and thus are described
as wildland fire spread models, while in some cases wildland fire spread is
used simply as a metaphor for some behaviour. These mathematical functions
include cellular automata and self-organised criticality, reaction-diffusion
equations, percolation, neural networks and others.  This section briefly
discusses some of the fire spread-related applications of these functions.

\subsection{Cellular automata and self-organised criticality}

Cellular automata (CA) are a formal mathematical idealisation of physical
systems in which space and time are discretised, and physical quantities take
on a finite set of values \citep{Wolfram1983}.  CA were first introduced by
Ulam and von Neumann in the late 1940s and have been known by a range of names,
including cellular spaces, finite state machines, tessellation automata,
homogenous structures, and cellular structures. CA can be described as discrete
space/time logical universes, each obeying their own local physics
\citep{Langton1990}.  Each cell of space is in one of a finite number of states
at any one time.  Generally CAs are implemented as a lattice (i.e. 2D) but can
be of any dimension.  A CA is specified in terms of the rules that define how
the state changes from one time step to the next and the rules are generally
specified as functions of the states of neighbours of each cell in the previous
time step. Neighbours can be defined as those cells that share boundaries,
vertices or even more further removed\footnote{In a 2D lattice, the cells
sharing boundaries form the von Neumann neighbourhood (4 neighbours), cells
sharing boundaries and vertices form the Moore neighbourhood (8 neighbours)
\citep{Albinet1986}}. The key attribute of a CA is that the rules that govern
the state of any one cell are simple and based primarily upon the states of its
neighbours, which can result in surprisingly complex behaviour, even with a
limited number of possible states \citep{Gardner1970} and can be capable of
Universal Computation \citep{Wolfram1986}.

Due to the inherent spatial capacity with interrelations with neighbouring
cells, CA have been used to model a number of natural phenomena, e.g. lattice
gases and crystal growth (with Ising models) \citep{Enting1977}, ecological
modelling \citep{Hogeweg1988} and have also been applied to the field of
wildland fire behaviour. \citet{Albinet1986} first introduced the concept of
fire propagation in the context of CA. It is a simple idealised isotropic model
of fire spread based on epidemic spread in which cells (or `trees') receive
heat from burning neighbours until ignition occurs and then proceed to
contribute heat to its unburnt neighbours. They showed that the successful
spread of the fire front was dependent upon a critical density of distribution
of burnable cells (i.e. `trees') and unburnable (or empty) cells, and that this
critical density reduced with the increasing number of neighbours allowed to
`heat' a cell. They also found that the fire front structure was fractal with a
dimension $\simeq$ 1.8. The isotropic condition in which spread is purely a
result of symmetrical neighbour interactions (i.e. wind or slope are not
considered) is classified as percolation (discussed below).
\citet{vonNiessen1988} extended the model of \citeauthor{Albinet1986} to
include anisotropic conditions such as wind and slope in which ignition of
crown and surface fuel layers was stochastic.

The idealised--more of a metaphor, really--`forest fire' model CA, along with
the sandpile (avalanche) and earthquakes models, was used as a primary example
of self-organised criticality \citep{Bak1987, Bak1996}, in which it is proposed
that dynamical systems with spatial degrees of freedom evolve into a critical
self-organised point that is scale invariant and robust to perturbation.  In
the case of the forest fire model, the isotropic model of \citet{Albinet1986}
was modified and investigated by numerous workers to explore this phenomenon,
eg. \citet{Bak1990, Chen1990, Drossel1992, Drossel1993, Clar1994, Drossel1996}
such that, in its simplest form, trees grow at a small fixed rate, \emph{p}, on
empty sites. At a rate \emph{f}, ($f<<p$), sites are hit by lightning strikes
(or matches are dropped) that starts a fire that burns if the site is occupied.
The fire spreads to every occupied site connected to that burning site and so
on. Burnt sites are then considered empty and can be re-colonised by new
growing trees. The rate of consumption of an occupied site is immediate, thus
the only relevant parameter is the ratio $\theta = p/f$, which sets the scale
for the average fire size (i.e. the number of trees burnt after one lightning
strike) (Grassberger 2002). Self-organised criticality occurs as a result of
the rate of tree growth and the size of fire that results when ignition
coincides with an occupied site, which in turn is a function of the rate of
lightning strike. At large $\theta$, less frequent but larger fires occur.  As
$\theta$ decreases, the fires occur more often but are smaller. This result
describes the principle underlying the philosophy of hazard reduction burning.
The frequency distribution of fire size against number of fires follows a power
law \citep{Malamud1998, Malamud1999} similar to the frequency distributions
found for sandpiles and earthquakes.

More recent work \citep{Schenk2000,Grassberger2002,Pruessner2002, Schenk2002}
has found that the original forest fire model does not truly represent critical
behaviour because it is not scale invariant; in larger lattices ($\simeq
65000$), scaling laws are required to correct the behaviour \citep{Pastor2000}.
\citet{Reed2002} compared size-distribution of actual burned areas in six
regions of North America and found that a simple power-law distribution was
`too simple to describe the distribution over the full range.
\citet{Rhodes1998} suggesting using the forest fire model as a model for the
dynamics of disease epidemics.

Self-organised criticality, however, is generally only applicable to the effect
of many fires over large landscapes over long periods of time, and provides no
information about the behaviour of individual bushfires. There are CA that have
used actual site state and neighbourhood rules for modelling fire spread but
these have been based on an overly simple understanding of bushfire behaviour
and their performance is questionable. \citet{Li2000, Li2003} attempted to
model the spread of individual bushfires across a landscape modelled in the 2D
CA lattice in which fuel is discrete and discontinuous.  While they supposedly
implemented the Rothermel wind speed/ROS function, their model shares more in
common with the Drossel-Schwabl model than any raster-based fire spread model.
\citeauthor{Li2000} determine critical `tree' densities for fire spread across
hypothetical landscapes with both slope and wind effects in order to study the
effect of varying environmental conditions on fire spread. However, ignition of
a cell or `tree' is probabilistic, based on `heat conditions' (or accumulated
heat load from burning neighbours); and the `tree' flammability is an arbitrary
figure used to differentiate between dead dry trees and green `fire-resistant'
trees. Essentially this is the same as tree immunity as proposed by
\citet{Drossel1993}. A critical density of around 41\% was found for lattices
up to 512 \ex 512 cells. The model is not compared to actual fire behaviour.

\citet{Duarte1997} developed a CA of fire spread that utilised a probabilistic
cell ignition model based on a moisture content driven extinction function
based \citet{Rothermel1972} using an idealised parameter based on fuel
characteristics, fuel moisture and heat load. Fuel was considered continuous
(but for differences in moisture). Duarte investigated the behaviour of the CA
and found the isotropic (windless) variant associated with undirected
percolation.  In the presence of wind, the CA belonged to the same universality
class (i.e. the broad descriptive category) as directed percolation. Duarte
notes that no CA at that time could explain the parabolic headfire shape
observed in experimental fires by workers such as \citet{Cheney1993}.

Rather than use hard and fast rules to define the states of a CA,
\citet{Mraz1999} used the concept of fuzzy logic to incorporate the descriptive
and uncertain knowledge of a system's behaviour obtained from the field in a 2D
CA. Fuzzy logic is a control system methodology based on an expert system in
which rates of change of output variables are given instead of absolute values
and was developed for systems in which input data is necessarily imprecise.
\citeauthor{Mraz1999} develop cell state rules (simple `if-then-else' rules)in
which input data (such as wind) is `fuzzified' and output states are
stochastic.

\citet{Hargrove2000} developed a probabilistic model of fire spread across a
heterogeneous landscape to simulate the ecological effects of large fires in
Yellowstone National Park, USA. Utilising a square lattice (each cell 50 m \ex
50 m), a stochastic model of fire spread in which the ignition of the Moore
neighbourhood around a burning cell is based upon an ignition probability that
is isotropic in no wind and biased in wind (using three classes of wind speed).
The authors determined a critical ignition probability (isotropic) of around
0.25 in order for a fire to have a 50\% chance of spreading across the lattice.
Spotting is modelled based on the maximum spotting distance as determined
within the SPOT module of the BEHAVE fire behaviour package \citep{Andrews1986}
using the three wind classes, a 3\degr  random angle from that of the
prevailing wind direction, and the moisture content of the fuel in the target
cell to determine spot fire ignition probability. Inclusion of spotting
dramatically increased the ROS of the fire and the total area burned.
Validation of the model, despite considerable historical weather and fire data,
has not been undertaken due to difficulties in parameterising the model and the
poor resolution of the historical data.

\citet{Muzy2002, Muzy2003, Muzy2005a, Muzy2005b}, \citet{Dunn2004} and
\citet{Ntaimo2004} explore the application of existing computational formalisms
in the construction of automata for the modeling of wildland fire spread;
\citet{Muzy2002,Muzy2003} and \citeauthor{Ntaimo2004} using Discrete Event
System Specification (DEVS or cell-DEVS) and \citet{Dunn2004} using CIRCAL.
DEVS attempts to capture the processes involved in spatial phenomena (such as
fire spread) using an event-based methodology in which a discrete event (such
as ignition) at a cell triggers a corresponding discrete process to occur in
that cell which may or may not interact with other cells. CIRCAL is derived
from a process algebra developed for electronic circuit integration and in this
case provides a rigorous formalism to describe the interactions between
concurrently communicating, interacting automata in a discretised landscape to
encode the spatial dynamics of fire spread.

\citet{Sullivan2004a} combined a simple 2-dimensional 3-state CA for fire
spread with a simplified semi-physical model of convection. This model explored
the possible interactions between a convection column, the wind field and the
fire to replicate the parabolic headfire shape observed in experimental
grassland fires \citep{Cheney1993}.  It used local cell-based spread rules that
incorporated semi-stochastic rules (allowing discontinuous, non-near neighbour
spread) with spread direction based on the vector summation of the mean wind
field vector and a vector from the cell to the centre of convection (as
determined by overall heat output of the fire as recorded in a six-stage
convection column above the fire). Fire shapes closely resembled those of fires
in open grassland but ROS was not investigated.

\subsection{Reaction-Diffusion}

Reaction-diffusion is a term used in chemistry to describe a process in which
two or more chemicals diffuse over a surface and react with one another at the
interface between the two chemicals. The reaction interface in many cases forms
a front which moves across the surface of the reactants and can be described
using travelling wave equations.  Reaction-Diffusion equations are considered
one of the most general mathematical models and may be applied to a wide
variety of phenomena.  A reaction-diffusion equation has two main components: a
reaction term that generates energy and a diffusion term in which the energy is
dissipated. The general solution of a reaction-diffusion equation is that of
the wave.

\citet{Watt1995} discussed the spatial dimensional reduction of a
two-dimensional reaction-diffusion equation describing a simple idealised fire
spread model down to a single spatial dimension. An analytical solution for the
temperature and thence the speed of the wave solution, which depends on the
reaction term and the upper surface cooling rate, is obtained from the reduced
reaction-diffusion equation by various algebra and linearisation.

\citet{Mendez1997} presents a mathematical analogue model of the cellular
automata `forest fire' model of \citeauthor{Drossel1992} in which they start
with a hyperbolic reaction-diffusion equation and then commence to apply
particular boundary conditions in order to determine the speed of propagation
of a front between unburnt or `green' trees on one side and `burnt' trees on
the other. It is assumed that in state 0 (all green) and state 1 (all burnt)
that the system is in equilibrium. Particular abstract constraints on the speed
of the front are determined and the nonequilibrium of the thermodynamics
between the two states are found.

\citet{Margerit1998} transformed the elliptical growth model of
\citet{Richards1990} in order to find an intrinsic expression for fire front
propagation.  The authors re-write the model in an optic geometric `variational
form' of the model in which the forest is represented as three-state `dots' and
the passage from unburnt or `at rest' dots and burning or `excited' dots is
represented by a wave front. This form is a Hamilton-Jacobi that when solved
gives the same result as Richards model. The authors then attempt to bring
physicality to Richards' model by proposing that the wave front is a
temperature isotherm of ignition. They put forward two forms: a hyperbolic
equation (double derivative Laplacian) and a parabolic (single derivative
Laplacian) which is the standard reaction diffusion equations (i.e. wave
solutions due to both the production of energy by a reaction and to the
transport of this energy by thermal diffusion and convection).  After some
particularly complicated algebra the authors bring the reaction-diffusion
equation back to an elliptical wave solution, showing that Richards' model is
actually a special case of the reaction-diffusion equation with geometry and no
physics.

\subsection{Diffusion Limited Aggregation}

\citet{Clarke1994} proposed a cellular automata model in which the key
propagation mechanism is that of diffusion limited aggregation (DLA). DLA is an
aggregation process used to explain the formation of crystals and snowflakes
and is a combination of the diffusion process with restriction upon the
direction of growth.  DLA is related to Brownian trees, self-similarity and the
generation of fractal shapes ($1 < D <2$). In this case, fire ignitions
(`firelets') are sent out from a fire source and survive by igniting new
unburnt fuel.  The direction of spread of the firelet is determined by the
combination of environmental factors (wind direction, slope, fuel).  If there
is no fuel at a location, the firelet `dies' and a new firelet is released from
the source.  This continues until no burnable fuel remains in direct connection
with the original fire source. Clarke's model is modified somewhat such that a
firelet can travel over fuel cells previously burned such that the fire
aggregates to the outer edge from the interior and wind, fuel and terrain are
used to weight or bias the direction of travel of the firelet.

The model was calibrated against an experimental fire conducted in 1986 that
reached 425 ha and was mapped using infrared remote sensing apparatus every 10
minutes. Environmental conditions from the experimental fire were held constant
in the model and the behaviour criteria of the cellular automata adjusted,
based on the spatial pattern of the experimental burn, its temperature
structure and temporal patten. Comparison of 100 fire simulations over the
duration of the experimental fire found that pixels that did actually burn were
predicted to burn about 80\% of the time. \citeauthor{Clarke1994} naively state
that fires burning across the landscape are fractal due to the self-similarity
of the fire perimeter but this ignores the fact that on a landscape scale, the
fire follows the fuel and it is the distribution of the fuel that may well be
fractal, not the fire as such.

A similar approach was taken by \citet{Achte2003} in which a ``rabbit'' acts
analogously to a fire, following a set of simple rules dictating behaviour,
e.g. rabbits eat food, rabbits jump, and rabbits reproduce. A hierarchy of
rules deal with rabbit death, terrain, weather, hazards.  An attempt to
incorporate rules regarding atmospheric feedbacks is also included. A similar
dendritic pattern of burning to that of \citet{Clarke1994} results when
conditions for spread are tenuous. Strong winds results in a parabolic headfire
shape.

\subsection{Percolation and fractals}

Percolation is a chemical and materials science process of transport of fluids
through porous materials. In mathematics it is a theory that is concerned with
transport through a randomly distributed media. If the medium is a regular
lattice, then percolation theory can be considered the general case of
isotropic CA.  Unlike CA, percolation can occur on the boundaries of the
lattice cells (bond percolation), as well as the cells themselves (site
percolation). Percolation theory has been used for the study of numerous
physical phenomena, from CO$_2$ gas bubbles in ice to electrical conductivity.
In addition to the CA models that have been applied to wildland fire behaviour
described above, several workers have investigated the application of
percolation itself to wildland fire behaviour.

\citet{Beer1990b,Beer1990c} investigated the application of isotropic
percolation theory to fire spread (without wind) by comparing predictions
against a series of laboratory experiments utilising matches (with ignitable
heads) placed randomly in a two-dimensional lattice. The intent was to simulate
the original percolation work of \citet{Albinet1986} using von Neumann
neighbours. They found that while the theory yielded qualitative information
about fire spread (e.g. that as the cell occupation density increased toward
the critical density, the time of active spread also increased) it was unable
to reproduce quantitatively the laboratory results. Effects such as radiant
heating from burning clusters of matches and convective narrowing of plumes
have no analogue in site/bond percolation where only near-neighbour
interactions are involved. They concluded that such models are unlikely to
accurately model actual wildfires and that models based on a two-dimensional
grid with nearest neighbour ignition rules are too naive.

\citet{Nahmias2000} conducted similar experimental work but on a much larger
scale, investigating the critical threshold of fire spread (both with and
without wind) across two-dimensional lattices. Utilising both scaled laboratory
experiments and field experiments in which the fuel had been manipulated to
replicate the combustible/noncombustible distribution of lattice cells in
percolation, \citeauthor{Nahmias2000} found critical behaviour in the spread of
fire across the lattice.  In the absence of wind, they found that the value of
the critical threshold to be the same as that of percolation theory when first
and second neighbours are considered. In the presence of wind, however, they
observed interactions to occur far beyond second nearest neighbours which were
impossible to predict or control, particularly where clusters of burning cells
were involved. They conclude that a simple directed percolation model is not
adequate to describe propagation under these conditions.

\citet{Ricotta2000,Caldarelli2001} investigated percolation in wildfires by
analysing satellite imagery of burned area (i.e. fire scars).  Both found the
final burned area of wildfires to be fractal\footnote{A fractal is a geometric
shape which is recursively self-similar (i.e. on all scales), defining an
associated length scale such that its dimension is not an integer, i.e.
fractional} (fractal dimension, $D_f \simeq 1.9$). \citet{Caldarelli2001} found
the accessible perimeter to have a fractal dimension of $\simeq 1.3$ and and to
be denser at their centres. They then show that such fractal shapes can be
generated using self-organised `dynamical' percolation in which a
time-dependent probability is used to determine ignition of neighbours.
Earlier, \citet{McAlpine1993} determined the fractal dimension of 14 fires (a
mix of data from unpublished fire reports and the literature) to be $\simeq
1.15$.  They then developed a method to convert perimeter predictions based on
elliptical length-to-breadth to a more accurate prediction using the fractal
dimension and a given measurement length scale.

\citet{Favier2004} developed an isotropic bond percolation model of fire spread
in which two time-related parameters, medium ignitability (the time needed for
the medium to ignite) and site combustibility (the time taken for the medium to
burn once ignited are the key controlling factors of fire spread. Favier
determined the critical values of these parameters and found fractal patterns
similar to that of \citet{Caldarelli2001}.

\subsection{Other methods}

Other mathematical methods that have been used to model the spread of wildland
fires include artificial neural networks \citep{McCormick1999}, in which a
large number of processing nodes (neurons) store information (weighting) in the
connections between the nodes.  The weightings for each connection between the
nodes is learnt by repeated comparing of input data to output data. Weightings
can be non-linear but necessarily needs a large dataset on which to learn and
assumes that the dataset is complete and comprehensive.  A related field of
endeavour is genetic algorithms \citep{Karaf1999, Karaf2004}, in which a pool
of potential algorithms are mutated (either at random or directed) and tested
against particular criteria (fitness) to determine the subset of algorithms
that will the next generation of mutating algorithms.  The process continues
until an algorithm that can carry out the specific task evolves.  This may lead
to local optimisations of the algorithm that fail to perform globally. Again,
the processes depends on a complete and comprehensive dataset on which to be
bred.

\Citet{Catchpole1989} modelled the spread of fire through a heterogeneous fuel
bed as a spatial Markov chain. A Markov chain is a time-discrete stochastic
process in which the previous states of a system are unrelated to the current
state of the system. Thus the probability of a system being in a certain state
is unrelated to the state it was in previously. In this context,
\citeauthor{Catchpole1989} treated, in one dimension, a heterogeneous fuel as a
series of discrete, internally homogeneous, fuel cells in which the variation
of fuel type is a Markov chain of a given transition matrix. The rate of spread
in each homogeneous fuel cell is constant related only to the fuel type of that
cell and to the spread rate of the cell previous.  The time for a fire to
travel through the \emph{n}th cell of the chain is then a conditional
probability based on the transition matrix. \citet{Taplin1993} noted that
spatial dependence of fuel types can greatly influence the variance of the rate
of spread thus predicted and expanded the original model to include the effect
of uncertainty in the spread rate of different fuel types.

The `forest-fire' model of self-organised criticality (discussed above) has led
to a variety of methods to investigate the behaviour of such models. These
include renormalisation group \citep{Loreto1995}, bifurcation analysis
\citep{Dercole2005}, and small-world networks \citep{Graham2003}.

\section{Discussion}

The field of computer simulation of fire spread is almost as old as the field
of fire spread modelling itself and certainly as old as the field of computer
simulation.  While the technology of computing has advanced considerably in
those years, the methods and approaches of simulating the spread of wildland
fire across the landscape has not changed significantly since the early days of
\citet{Kourtz1971} on a CDC6400 computer.  What has changed significantly has
been the access to geographic data and level of complexity that can be
undertaken computationally. The two areas of research covered in this paper,
computer simulation of fire spread and the application of mathematical
analogues to fire spread modelling are very closely related.  So much so that
key methods can be found in both approaches (e.g. raster modelling and cellular
automata (percolation)).

The discussion of simulation techniques concentrated on the various methods of
converting existing point (or one dimensional) models of rate of forward spread
of a fire to two dimensions across a landscape. The most widely used method is
that of Huygens' wavelet principle, which has been used in both vector (line)
and raster (cell) simulation models.  The critical aspect of this method is the
choice of a template shape for the spawned wavelet (or firelet).  The most
common is that of the simple ellipse but \citet{Richards1995a} showed that
other shapes are also applicable.

\citet{French1992} found that vector-based simulations produce a much more
realistic fire perimeter, particularly in off-axis spread, than raster-based
simulations. However, raster-based simulations are more proficient at dealing
with heterogeneous fuels than vector-based models.  Historically, the
requirements of raster-based models to have raster geographic data at a high
enough resolution to obtain meaningful simulations has meant that vector-based
models have been favoured, but the reducing cost of computer storage has seen a
swing in favour of raster-based models.  Increasingly, the choice of type of
fire spread model is driven by the geographic information system (GIS) platform
in which the geographic data resides, making the decision between the two moot.

Alternative fire propagation methods such as \citet{Clark2004} are restricted
due to the specificity of the model in which the method is embedded.  The
coupled fire-atmosphere model is unique in the field of fire spread simulation
in that it links a fully formed 3D mesoscale meteorological model of the
atmosphere with a 1D fire spread model.  In order to make the simulation work,
the authors had to devise methods of propagating the fire perimeter at a
resolution much smaller than the smallest resolvable area of the coupled model.
The result is a method specific to that model and one which simply replaces the
broad generalised approach of the template ellipse with a much more fundamental
(but not necessarily more correct) variable spread direction around the
perimeter.

The broad research area of mathematics has yielded many mathematical functions
that could be seen as possible analogies to the spread of fire across the
landscape.  The most prevalent of these is the cellular automata model (or the
much more general percolation theory).  These models are well suited to spatial
propagation of an entity across a medium and thus have found application in
modelling epidemic disease spread, crowd motion, traffic and fire spread across
the landscape.  The approaches taken in such modelling range from attempting to
model the thermodynamic balance of energy in combusting cells to treating the
fire as a contagion with no inherent fire behaviour whatsoever apart from a
moving perimeter. The later models have found great use in the exploration of
critical behaviour and self-organisation.

Other approaches, such as reaction-diffusion models, have had a more physical
basis to the modelling of fire spread but their application to fire assumes
that fire behaviour is essentially a spatially continuous process that does not
include any discontinuous processes such as spotting. Models based on this
mathematical conceit generally have to have specific components fitted to the
original model to incorporate fundamental combustion processes such as
spotting, non-local radiation, convection, etc.

While the mathematical analogue models discussed here may appear to have little
in conjunction with real fire behaviour, they do offer a different perspective
for investigating wildland fire behaviour, and in many cases are more
computationally feasible in better than real time than many physical or
quasi-physical models. Divorced as they are from the real world, approaches
such as that of percolation theory and cellular automata provide a reasonable
platform for the biological heterogeneity of fuels across the landscape.  This
is confirmed by the fact that many GIS platforms take such a raster view of the
world. However, to simulate the spread of a fire across this (possibly 3D)
surface, requires the development of rules of propagation that incorporate both
local (traditional CA) rules with larger scale global rules in order to
replicate the physical local and non-local processes involved in fire spread.
The foremost of these is the importance of the convection column above a fire.
The perimeter of a fire, although it may seem to be a loosely linked collection
of independently burning segments of fuel, actually does behave has a
continuous process in which the behaviour of the neighbouring fuel elements
does affect the behaviour of any single fuel element.  Other non-local
interactions include spotting and convective and radiative heating of unburnt
fuels.

Regardless of the method of fire propagation simulation that is used, the
underlying fire spread model that determines the behaviour of the simulation is
the critical element.  The preceding two papers in this series discussed the
various methods, from purely physical to quasi-physical to quasi-empirical and
purely empirical, that have been developed since 1990.  It is the veracity,
verification and validation of these models that will dictate the quality of
any fire spread simulation technique that is used to implement them.  However,
performance of the most accurate fire spread model will be limited, when
applied in a simulation, to the quality of the input data required to run it.
The greater the precision of a fire spread model, the greater the precision of
the geographical, topographical, meteorological and fuel data needed to achieve
that precision--one does not want a model's prediction to fail simply because a
there was a tree beside a road that was not present in the input data! The need
for greater and greater precision in fire spread models should be mitigated to
a certain extent by the requirements of the end purpose and quantification of
the range of errors to be expected in any prediction.

As in the original attempts at `simulation' using a single estimate of forward
spread rate and a wall map to plot likely head fire locations, in many
instances highly precise predictions of fire spread are just not warranted,
considering the cost of the prediction and access to suitable high quality
data. For the most part, `ball-park' predictions of fire spread and behaviour
(not unlike those found in current operational prediction systems) are more
than enough for most purposes. An understanding of the range of errors, not
only in the input data but also in the prediction itself, will perhaps be more
efficient and effective use of resources.

In the end, there will be a point reached at which the computational and data
cost of increases in prediction accuracy and precision will outweigh the cost
of cost-effective suppression.

\section{Acknowledgements}

I would like to acknowledge Ensis Bushfire Research and the CSIRO Centre for
Complex Systems Science for supporting this project; Jim Gould and Rowena Ball
for comments on the draft manuscript; and the members of Ensis Bushfire
Research who ably assisted in the refereeing process, namely Miguel Cruz and
Ian Knight.

\bibliographystyle{apalike}

\end{document}